# Fabrication of ultra-smooth, high-aspect ratio, sub-10 nanometer nanostructures

John A. Scott,[*,†] Panaiot G. Zotev,[‡] Luca Sortino,[¶] Yadong Wang,[‡] Amos Akande,[†] Michelle L. Wood,[§] Alexander I. Tartakovskii,[‡] Milos Toth,[*,‖,⊥] and Stefano Palomba[*,†]

[†]Institute for Photonics and Optical Sciences (IPOS), School of Physics, The University of Sydney, Camperdown, NSW, 2006, Australia

[‡]School of Mathematical and Physical Sciences, University of Sheffield, Sheffield S3 7RH, United Kingdom

[¶]Chair in Hybrid Nanosystems, Nanoinstitute Munich, Faculty of Physics, Ludwig-Maximilians-Universität München, 80539 Munich, Germany

[§]Sydney Analytical, The University of Sydney, Camperdown, NSW, 2006, Australia [‖]Mathematical and Physical Sciences, University of Technology Sydney, Ultimo, NSW, 2007, Australia

[⊥]ARC Centre of Excellence for Transformative Meta-Optical Systems, University of Technology Sydney, Ultimo, NSW, 2007, Australia

E-mail: John.Scott1@sydney.edu.au; Milos.Toth@uts.edu.au; Stefano.Palomba@sydney.edu.au

**Abstract**

Deterministic and versatile approaches to sample preparation on nanoscopic scales are important in many fields including photonics, electronics, biology and material science. However, challenges exist in meeting many nanostructuring demands—particularly in emerging optical materials and component architectures. Here, we report a nanofabrication workflow that overcomes long-standing challenges in deterministic and top- down sample preparation procedures. The salient feature is a carbon mask with a low sputter yield that can be readily shaped using high resolution electron beam processing techniques. When combined with focused ion beam processing, the masking technique yields structures with ultra-smooth, near-vertical side walls. We target different material platforms to showcase the broad utility of the technique. As a first test case, we prepared nanometric gaps in evaporated Au. Gap widths of 7 ± 2 nm, aspect ratios of 17, and line edge roughness values of $3\sigma$ = 2.04 nm are achieved. Furthermore, the gap widths represent an order of magnitude improvement on system resolution limits. As a second test case, we designed and fabricated dielectric resonators in the ternary compounds $MnPSe_3$ and $NiPS_3$; a class of van der Waals material resistant to chemical etch approaches. Nanoantenna arrays with incrementally increasing diameter were fabricated in crystalline, exfoliated flakes. The optical response was measured by dark field spectroscopy and is in agreement with simulations. The workflow reported here leverages established techniques in material processing without the need for custom or specialized hardware. It is broadly applicable to functional materials and devices, and extends high speed focused ion beam milling to true sub-10 nm length scales.

## Introduction

The preparation of nanostructures is integral to the development of many new technologies across diverse fields. Among the different approaches, lithographic methods have been enormously successful in both commercial (nano)manufacturing, including integrated complementary metal oxide semiconductor (CMOS) devices, and in process development across different research fields[1,2,3,4]. Despite this success, lithographic-based approaches face numerous challenges in meeting the demands of emerging materials, component architectures, non-planar substrates, cost-efficient prototyping, and processing on sub-10 nm length scales[5,6].

Drawbacks include the line edge and sidewall surface roughness, material specificity, poor mask etch selectivity, and spatial resolution limits – particularly in high-aspect ratio nanostructures. These limiting factors are particularly detrimental to the performance and reproducibility of light-interacting nanostructures with engineered optical responses[7,8,5,9]. This has spawned interest in alternative and complimentary nanofabrication approaches including focused ion beam (FIB) machining methods[10,11,12,13,14,15,16].

Focused ion beam systems provide high resolution, site-selective ion processing of materials[17]. Increasingly, they have been utilized in the fabrication and editing of active components for nanoscale technologies, exemplified by their pivotal role in recent research on metasurfaces (alongside lithographic based processes)[18,19]. The deployment of focused ion beam systems is largely facilitated by high resolution, direct-write capabilities that are material agnostic. Efforts to expand the utility of FIB systems—in both traditional use cases and beyond—have focused on inherent (system) shortcomings. These efforts include the development of reactive and inert plasma ion sources, improved processing throughput (i.e. greater accessible beam currents), damage mitigation, preservation of material stoichiometry during ion processing and FIB probe resolution[20,21,22,23,24]. However, despite impressive solutions implemented to drive the evolution of FIB instruments, challenges still remain. In particular, editing, fabrication and sampling of functional nanostructures, which underpin many emerging technologies.

Here we report a nanofabrication workflow involving a low sputter yield carbon deposit mask and focused ion beam irradiation. By leveraging high resolution electron beam processing steps (deposition and etching), we are able to directly-write masks with high etch selectivity and exceptional sidewall smoothness. To demonstrate the broad utility of the workflow, we prepared optically-active nanostructures in different material systems and confirmed their functionality by measuring their optical response. The structures we targeted are of technological significance and their fabrication is hindered or prohibited by shortcomings of existing techniques. As a first test case, we prepared high-aspect ratio, nanometric gaps in evaporated gold (200 nm thick). Gap widths as narrow as 7 ± 2 nm, aspect ratios of 17 and line edge roughness (3$\sigma$) values of 2.04 nm were achieved. The demonstrated gap width is an order of magnitude improvement on the intrinsic resolution limit of the employed plasma focused ion beam system. Moreover, the fabricated gap profile is superior to competing techniques, of which ultra-high-resolution He ion beam sputtering is benchmarked here. We used surface enhanced Raman spectroscopy of the analyte *crystal violet* to confirm the optical response of the fabricated nanostructures.

As a second test case we prepared dielectric nanoantennas in the ternary compounds $MnPSe_3$ and $NiPS_3$. The van der Waals crystals are an emerging class of magnetic dielectrics that have garnered interest through their potential for spin correlated physics—combining high optical confinement and pronounced magneto-excitonic coupling[25,26]. Yet, to date, engineering of the electromagnetic response through nanostructuring has not been achieved. Materials of complex chemical composition—including ternary compounds—are resistant to chemical 'dry' etch nanofabrication approaches. This is due to the necessity of forming volatile reaction products with each constituent element. Physical, sputtering based etch approaches using polymer masks overcome this limitation. However they suffer from numerous problems including redeposition and resist mask etch selectivity, for example selectivities of ~1 are reported for $LiNbO_3$[27]. Here we prepared nanoantenna arrays in exfoliated $MnPSe_3$ and $NiPS_3$ crystalline flakes. We demonstrate nanoscale tunability by preparing antennas with incrementally increasing diameter. We show excellent uniformity and reproducibility across multiple arrays and confirm the presence of Mie resonances by dark field spectroscopy. Our work shows a nanofabrication procedure that overcomes limitations in common techniques. The workflow has the potential for broad applicability in sub-10 nm editing, fabrication and characterization.

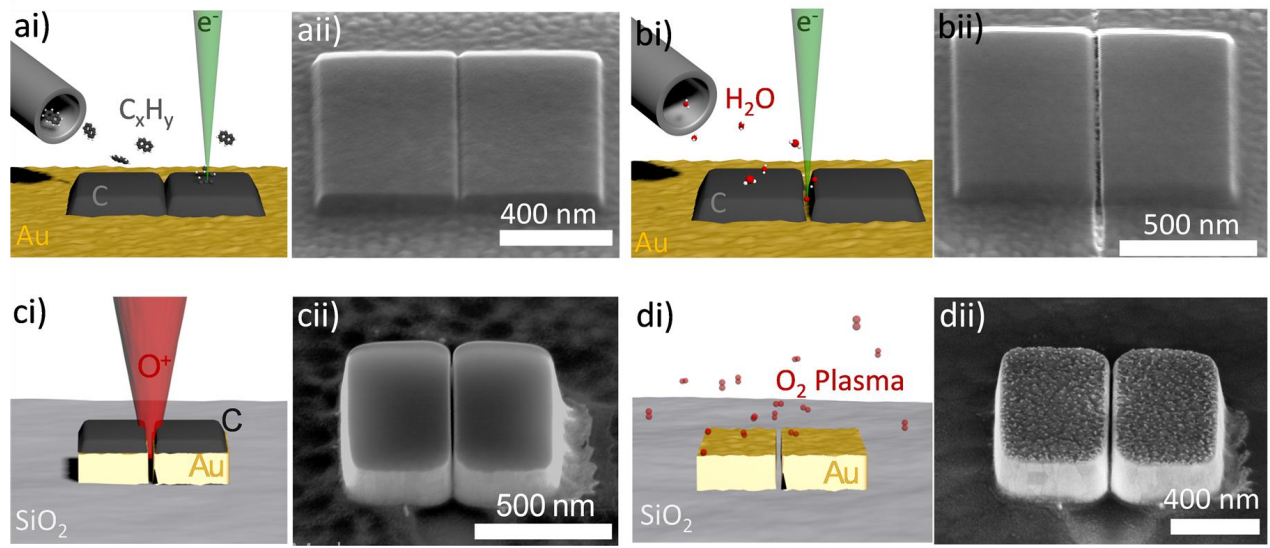

Figure 1: **Carbon masking nanofabrication workflow. (a)** Schematic representation of focused electron beam induced deposition (FEBID) of carbon deposits on Au and (ii) a tilted- SEM image of the rectangular prism FEBID C deposits. **(b)** Schematic showing the focused electron beam induced etching step used to remove excess C in the gap region between adjacent rectangular prisms and (ii) a SEM image of the resulting etch. **(c)** Schematic showing focused ion beam milling of Au dimer structures with C deposit masking and (ii) a SEM image of the structures following ion beam milling. **(d)** Schematic and SEM image following oxygen plasma processing step for carbon mask removal.

## Results and discussion

### Workflow overview

We start with a general overview of the steps involved with the workflow. These steps are illustrated in Figure 1 and include: (i) focused electron beam deposition of a low sputter yield C mask (Figure 1a), (ii) removal of unwanted C using a focused electron beam facilitated etch step (Figure 1b), (iii) focused ion beam milling (Figure 1c) and, finally, (iv) removal of the C mask using an oxygen plasma (Figure 1d). A detailed description of the fabrication process is provided in the Methods section. The experimental work in steps (i) - (iii) was performed on a multi-ion source, plasma FIB dual beam. This instrument was employed for two main reasons: (i) the ability to select the primary ion species is important for the work- flow (discussed in greater detail below), and (ii) plasma FIB systems trade off high process throughput (i.e., high beam current) for poor lateral resolution (i.e., a large ion beam diam- eter)[21]. This trade-off can be inverted (by, for example, using a He ion beam instrument). However, we instead achieve ultra-high spatial resolution using a high-throughput plasma FIB. This combination of resolution and throughput is important for scalable high-resolution nanofabrication, and for broadening the utility of ion beam techniques.

The first steps of the workflow involve the creation and shaping of a carbon mask (Figure 1a). Deposition of a carbonaceous material was performed by focused electron beam induced deposition (FEBID). Here, a carbon-containing gaseous precursor was introduced at the sample surface using a commercial gas injection system. Under focused electron irradiation, the precursor is dissociated, leaving behind non-volatile constituents to form a deposit. The composition consists of hydrogenated disordered carbon with a mixture of $sp^2$ and $sp^3$ bonding states[28]. Electron beam parameters and scan strategies are used to influence the 3-dimensional shape of the deposit. A schematic representation of the technique is shown in Figure 1a(i) as well as a tilted SEM image of a typical deposit consisting of adjacent rectangular prisms. The important properties of the C deposits are their very low sputter yield and their smooth sidewalls. The low sputter yield is due to the bonding structure (and density), consistent with similar reports[29]. A relatively high electron beam energy was chosen for improved side-wall angle at the cost of the deposition rate. We note that whilst the separation distance between the rectangular patterns was defined as 40 nm in the pat- tern, the gap width narrows with deposition time. We attribute this proximity effect to the high secondary electron yields at the deposit sidewalls causing precursor dissociation and scattering of primary electrons by gas phase species. This process is known as beam skirting and is evidenced by a halo like, thin layer of material around the beam patterned area[29]. An annotated SEM image showing the electron beam pattern and the resulting deposition halo is given in Figure S1 (Supporting Information).

Next, an electron beam facilitated etch step was performed. This step was used to remove the unwanted C that has built up in the gap region between the deposits, though, more generally, it may be used to make high resolution adjustments to the C mask[30,31,32]. Here, we removed the unwanted C by focused electron beam irradiation in the presence of $H_2O$. A critical aspect of this step is that a *volatile* reaction product forms between the dissociated precursor and the material. In the case of carbon-based materials, $H_2O$ is a well established precursor for high resolution chemical etching, including in materials such as diamond[30,31]. A schematic of the process, as well as a tilted SEM image following the processing step are shown in Figures 1b(i) and (ii), respectively. Comparing SEM images before (Figure 1a(ii)) and after etching (Figure 1b(ii)) it is clear that removal of C has occurred, restoring a gap between the adjacent rectangle masks. If this step was not performed, complete masking of the focused ion beam occurs and milling of the underlying Au layer was prevented. This is shown in Figure S2.

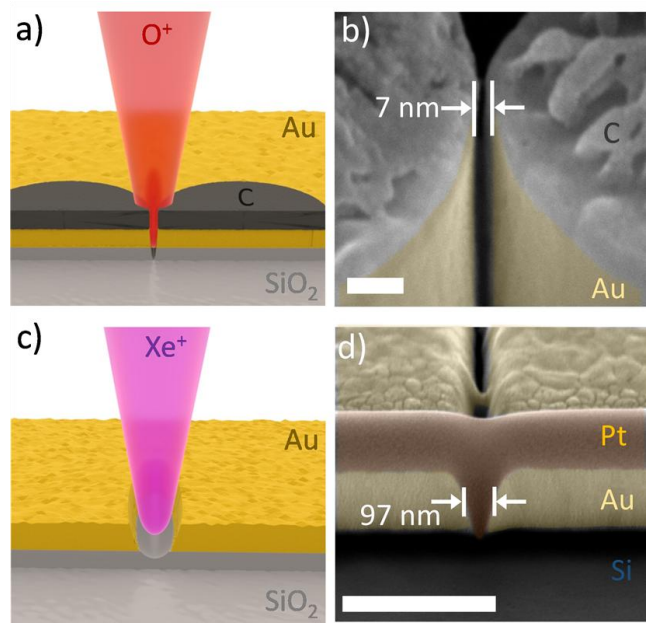

Figure 2: **FIB mill gap width with and without carbon masking workflow procedure. (a)** Schematic showing cross section of FIB mill in the presence of a C deposit mask.
**(b)** SEM image of FIB milled gap width in Au (in the presence of C deposit masking).
**(c)** Schematic showing cross section of FIB mill in the absence of a C deposit mask. **(d)** SEM image of FIB milled gap in Au (in the absence of C deposit masking). The scale bars represent 40 nm and 500 nm in (b) and (d), respectively. SEM images are falsely colored to highlight each elemental layer.

The next step, following mask preparation, was removal of the unmasked gold (Au)

by focused ion beam milling (Figure 1c). This was performed in a typical fashion, with one exception. Here, oxygen was selected as the primary ion species as it allows for the enhanced removal of C through a process known as *chemical sputtering* [33]. Normally enhanced removal of the mask is undesirable, here, it was determined to be necessary. When control experiments were performed using an inert Xe⁺ focused ion beam, C was observed to build up in the gap region (Figure S3). The process that results in this is known as re-deposition and is a common occurrence during physical sputtering. Oxygen, on the other-hand, acts to prevent this through enhanced C removal by chemical processes. Figure 1c(i) shows a schematic of oxygen focused ion beam irradiation producing adjacent Au rectangular prisms and Figure 1c(ii) shows a SEM image after the focused ion beam processing step. Finally, the sample was removed and the carbon mask was etched in an oxygen plasma asher before imaging (Figures 1d(i) and (ii)). Some carbonaceous residue remained on the surface after this step, this is due to the chemically resilient bonding structure, likely sp$^3$ carbon. Each step deployed in this workflow is common to many established sample preparation procedures and no specialized, nonstandard hardware was used.

Notably, measurements of the line edge roughness (LER) were performed on the adjacent rectangular prism structures (Figure S4). Here we determined the LER to be $3\sigma$ = 2.04 nm. The low roughness value is attributed to smooth mask sidewalls and accelerated ion beam polishing effects. The latter is analogous to surface polishing by glancing incidence FIB milling[34]. Polishing of this nature overcomes roughening due to grains in deposited metal films as well as undesirable masking effects. This is particularly beneficial for low loss plasmonic devices which rely on large grain sizes (hence a large potential line edge roughness) to reduce electron scattering[35].

Next we show a side-by-side comparison of the milled gap width, *with* and *without*, the C masking workflow. The important feature is the low sputter yield C mask that efficiently blocks part of the focused ion beam and prevents milling of the underlying layer (Figure 2a). Through the high resolution techniques used to deposit and shape the mask, super-resolution focused ion beam milling was achieved. This is shown in Figure 2b by the milling of a 7 ± 2 nm gap in an evaporated Au film. Additional high magnification SEM characterization can be found in the Supporting Information Figure S5. In the absence of the mask (Figure 2c) the gap width was 97 nm (FWHM) (Figure 2d). Here Pt was deposited as a capping layer for cross-sectioning of the line mill. In this reference control experiment (Figure 2d) the beam parameters (including the ion species Xe$^+$) were selected as representative high resolution beam parameters. In other words, the best conditions for producing narrow gap widths. The measured width is consistent with reported values[20].

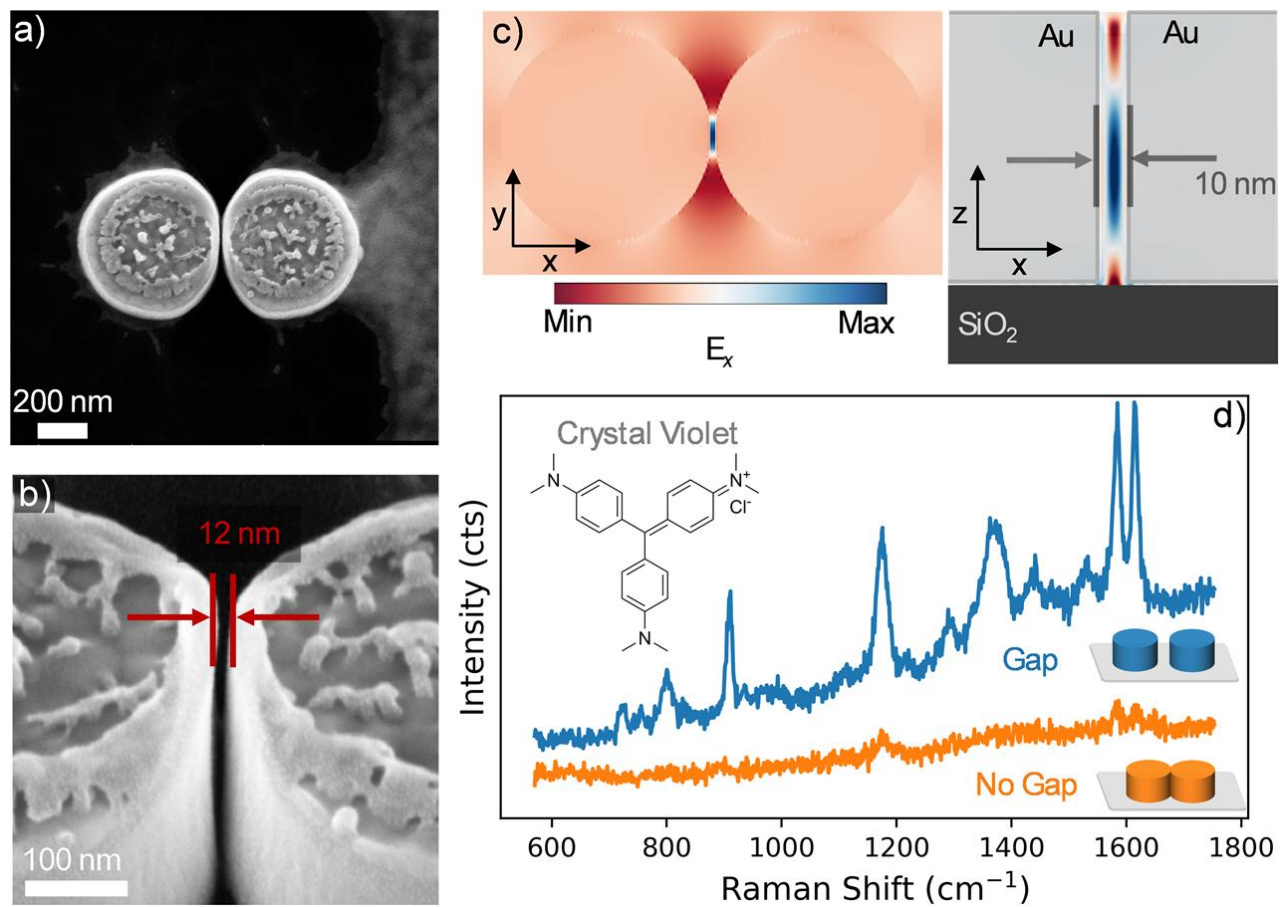

Figure 3: **Surface Enhance Raman Spectroscopy. (a)** Top down and **(b)** tilted SEM image of Au dimer structure. **(c)** 2D slices along the XY and XZ planes of 3D FDTD simulation results showing electric field intensity in the gap region (polarization of incident plane wave was set along the x-axis). **(d)** Crystal violet Raman spectra acquired from an Au dimer structure (blue) and an Au structure with no gap (orange).

In contrast to inductively coupled plasma source FIB systems, gas field-ionization source

(GFIS) helium ion microscopes (HIMs) are dedicated high resolution tools that are used routinely in the preparation of plasmonic gap structures. The ion source and column are designed to generate probe diameters down to 0.35 nm[24]. However, when used to produce high-aspect ratio gap structures, the GFIS beam profile results in a mill with a Gaussian- like shape. We demonstrate this by performing a single pixel wide line mill pattern in an evaporated Au film on $SiO_2$. The Au layer is identical to that used in Figure 2. A gap distance of ~50 nm was measured at the gold surface, that quickly narrows to 6 nm approximately halfway through the 200 nm thick Au film (Figure S6). Tapering of this nature supports optical modes that may not be desirable for the intended application, yet remain present due to uncontrollable fabrication limitations.[36,37].

In addition to the tapering mill profile and low processing throughput, an unintended and undesirable consequence of the He ion source is material swelling. This has the effect of limiting the ion dose density and the materials (including substrates) that can undergo editing and fabrication. A phenomenon that is much less consequential in plasma FIB systems on account of the reduced implantation depth of $O^+$ and $Xe^+$ ions. In Figure S7 we show the effect of swelling when a silicon substrate is used to support a 200 nm Au film. Here we performed comparable He ion mills starting with the same parameters used for the high-resolution line mill (0.5 nC/$\mu$m). The swelling in silicon results in a localized bulge in the Au surface with a height of ~100 nm. The swelling behavior was observed to be dose dependent by performing a series of line mills with the following parameters: 0.5, 0.6 and 0.7 nA/$\mu$m and its severity prevented the fabrication of functional, high-aspect ratio, gap structures.

## Plasmonic dimer resonator

To demonstrate the broad utility of the nanofabrication workflow, we prepared nanostructures in different materials and confirmed their optical response. We started with plasmonic dimer structures in an evaporated Au film. The preparation of high-aspect ratio, ~10 nm gap structures is important for many plasmonic technologies including integrated planar hybrid devices[38,39,40]. Plasmonic dimer structures for surface enhanced Raman scattering (SERS) measurements, were prepared from circular C masks (500 nm in diameter) deposited on a 200 nm thick Au layer. After focused ion beam milling and removal of the C mask in the O₂ asher (see Methods for details), SEM characterization of the structures was performed. High resolution imaging revealed the presence of a 12 nm gap (Figures 3a and b). To confirm high optical confinement, and field enhancement in the gap region, 3D FDTD simulations were performed using MEEP[41]. Figure 3c shows the electric field intensity in the gap region resulting from an incident plane wave source with a vacuum wavelength of 532 nm. The polarization was set along the x-axis. After fabrication, a dilute (24 µmol) solution containing the molecule *crystal violet* was dropcast onto the substrate to investigate SERS enhancement of the dimer structure. Figure 3d shows Raman spectra acquired from a dimer structure with a gap (blue trace) and a similar structure with no gap (orange trace). We observe the strongest Raman signal intensity from the gap dimer structure and the spectrum clearly resolves Raman peaks that are absent in the reference control structure (Figure S8). The peaks were confirmed to match that of crystal violet by measuring a high concentration (2 mM) solution, dropcast onto stainless steel (Figure S9).

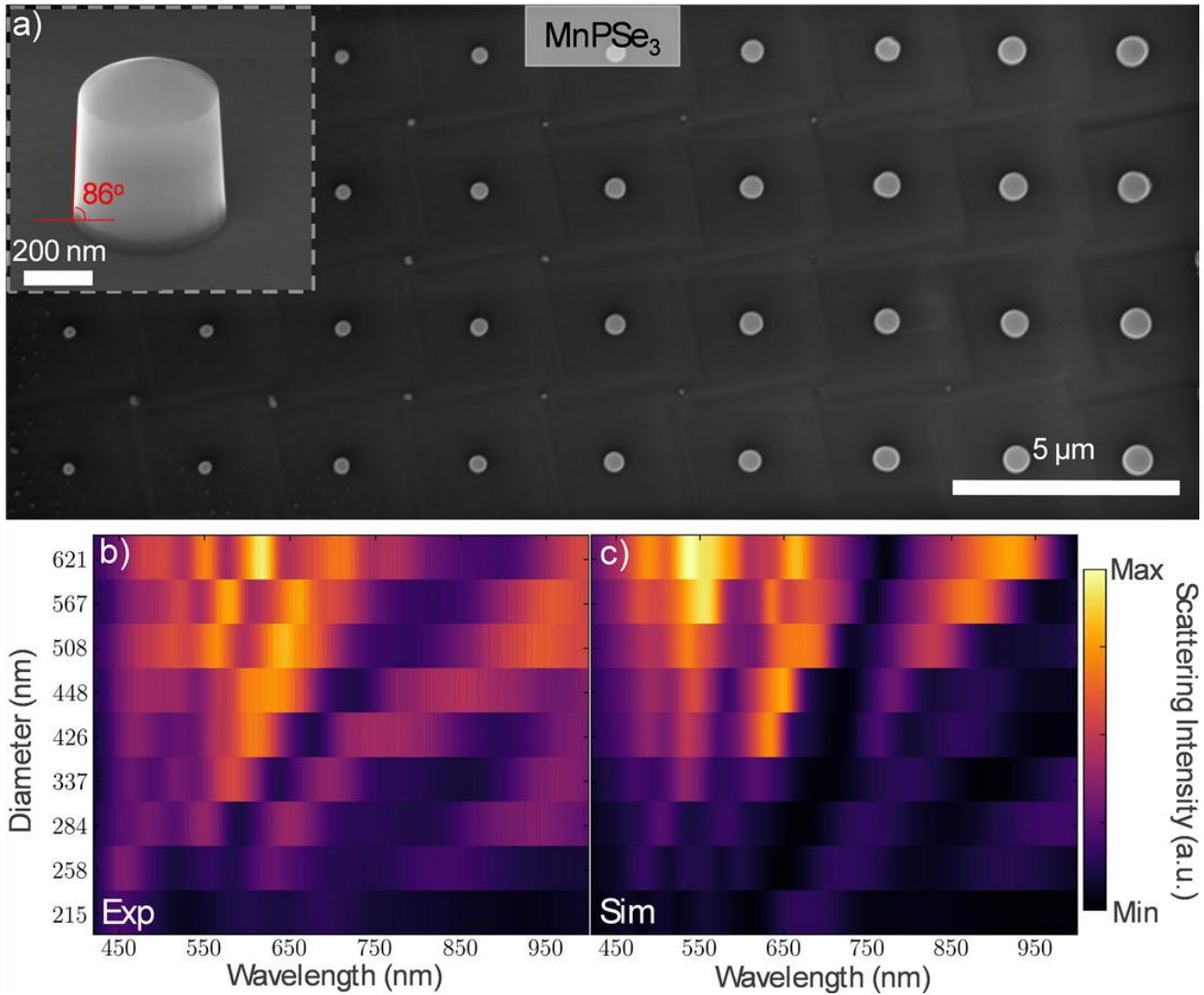

Figure 4: **MnPSe₃ nanoantenna array and optical characterization (a)** SEM image of MnPSe$_3$ resonator array including a high magnification, tilted SEM image of an individual structure (inset). (b) Heat map plot showing dark field scattering spectra of nanoantennas with increasing diameter and (c) simulated dark field scattering specra for nanoantennas.

**Dielectric nanoantennas**

Next we show that the workflow can be used more generally for the preparation of nanostructured materials. Dielectric resonant nanostructures (including nanoantennas) have demonstrated unrivaled versatility in the manipulation of light. This burgeoning field within nanophotonics provides a new paradigm of sub-wavelength light-matter interactions that has applications in communications, computation, sensing and quantum technologies[19,42]. Artificial engineering of the electromagnetic resonances underpins development within the field. Rapid and versatile methods of sample preparation, compatible with hybrid mate- rial/component integration, has the potential to greatly drive advancements. Particularly for operation at terahertz frequencies and higher, where few fabrication approaches are able to meet the resolution demands[42].

Here we target nanoantenna arrays in the van der Waals crystals $MnPSe_3$ and $NiPSe_3$. We do this for several reasons. Firstly ternary compound materials are resistant to fabrication involving 'dry' plasma chemical etch processes, due to the complexity in their chemical make- up[25]. Ultimately ion processing may involve approaches incorporating physical sputtering, in which case, suitable masking is required, of which diamond-like ($sp^3$ bonded) carbon has emerged as a viable candidate[43]. Traditional resist-based masking approaches suffer from poor etch selectivity. In the case of LiNbO$_3$, the mask etch selectivity is ~1 for common resists, whilst in the case of diamond-like C films selectivities of 3 have been reported[27,43]. Here we achieved selectivities of 3.3 and 2.5 for MnPSe$_3$ and NiPS$_3$ respectively (based on Monte Carlo simulation techniques using SRIM)[44]. Secondly, one of the salient feature of the workflow is the ability to produce near vertical, smooth sidewalls. This is a feature that is important to photonic structures including components/meta-atoms in metasurfaces[7,8,45].

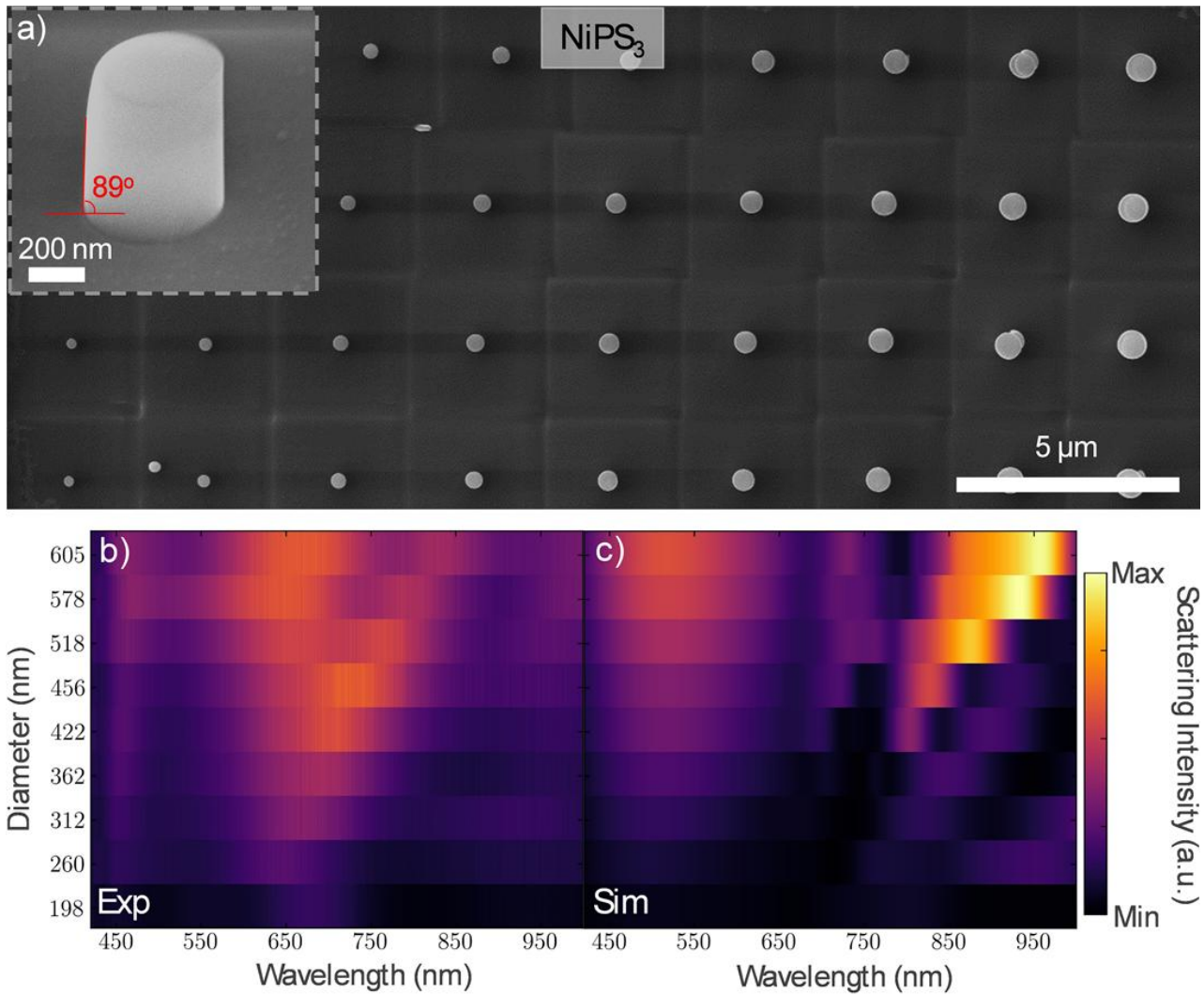

Figure 5: **NiPS$_3$ nanoantenna array and optical characterization.** (a) SEM image of NiPS$_3$ resonator array including a high magnification, tilted SEM image of an individual structure (inset). (b) Heat map plot showing dark field scattering spectra of nanoantennas with increasing diameter and (c) simulated dark field scattering specra for nanoantennas.

Lastly, focused electron beam induced deposition is a high resolution, direct-write technique. It offers great flexibility in mask preparation and is compatible with hybrid material integrated devices. We illustrate the former point by preparing nanoscale antennas with incrementally increasing diameters that demonstrate a high degree of uniformity in antennas with the same design.

We first identified flakes of suitable thickness (~400 nm) (Figure S10 and S12) from mechanically exfoliated $MnPSe_3$ and $NiPS_3$ crystals. Patterning C FEBID masks was then performed. FEBID masks starting from 200 nm in diameter and increasing in increments of 50 nm to a maximum of 600 nm were deposited. Next FIB irradiation was performed using a $Xe^+$ beam (full procedure is detailed in the Methods section). Here gap structures were not being prepared therefore oxygen primary ions were avoided. This highlights a modified approach that may be used for oxygen sensitive materials and mitigates the influence of implanted ions (a major limitation of He and Ga ion microscope systems). Characterization of the nanoantenna array after fabrication is shown in Figure 4a. From the tilted SEM image (Inset Figure 4a) the side walls showed negligible roughness with an angle of $86^o$. Figure 4b-c show a 2D heat map plot of the dark field spectra and the corresponding FDTD simulations, for a single row of nanoantennas. From the measured response we observe the emergence of optical Mie resonances, and an increased scattered cross section with increasing diameter. Importantly, we observe a clear dispersion relationship, with a red-shift of the resonance peaks with increasing structure diameter, which follows the expected behavior of dipolar and multipolar resonances in dielectric nanoantennas. Figure S11 shows the dark field spectra for each row in the array and confirms the excellent reproducibility between arrays. The same procedure was performed on a 390 nm thick $NiPS_3$ flake. The results are shown in Figure 5 and Figure S13.

## Conclusion

We demonstrate a fabrication workflow that addresses limitations in existing techniques. The workflow leverages high-resolution electron beam processing steps to deposit and shape very low sputter yield carbon masks. When utilized, in conjunction with focused ion beam processing, facilitates super-resolution ion beam processing on sub-10 nm length scales and yields nanostructures with ultra-smooth morphology. We demonstrate the broad utility of the nanomanufacturing protocol using 2 sets of examples and confirm their optical resonances. Firstly we prepared high-aspect ratio, nanometric gaps in evaporated Au films. Gap distances of 7 ± 2 nm, aspect ratios of 17 and line edge roughness (3$\sigma$) values of 2.04 nm are achieved. The gaps support surface enhanced Raman spectroscopy of a target analyte. Secondly, we prepared dielectric nanoantenna arrays in the van der Waals crystals: $MnPSe_3$ and $NiPS_3$. We achieve smooth, near vertical sidewalls with excellent uniformity across the diameter range confirmed by dark field spectroscopy. The workflow enables rapid prototyping of functional nanostructures and has the potential for broad applicability in sub-10 nm editing, fabrication and characterization.

## Experimental

**Carbon masking workflow Au structures**: Starting with a 200 nm evaporated Au film on Si/$SiO_2$, a hard C mask was fabricated by focused electron beam induced deposition (FEBID). C FEBID mask deposits were produced here using a 20 keV, 800 pA electron beam, and the precursor species: naphthalene ($C_{10}H_8$). The dwell time was 1 $\mu$s. The rectangular pattern dimensions are 1 $\mu$m x 500 nm, with a separation distance of 40 nm (Figures 1 & S1-S3). The total electron beam irradiation time was 2 minutes per pattern or 4 minutes in total (Figure 1). However, we note that the deposition time changed depending on precursor depletion in the crucible. In some instances the total time was 3 minutes per pattern or 6 minutes total deposition time (Figures S1-S3). Circular pattern dimensions were 500 nm diameter and a separation of 40 nm (Figures 2 and 3) and the patterning time was 3 minutes per structure (6 minutes in total). Patterning of the dimer structures was performed in a parallel fashion (as apposed to serial) to achieve uniform deposit heights. The chamber pressure for deposition was ~5 x $10^{-5}$ mbar increasing from a system base pressure of 4 x $10^{-6}$.

Focused electron beam induced etching (FEBIE) was performed in a similar fashion. The $H_2O$ precursor is housed in a seperate crucible in the same gas injection system and uses common delivery lines. It was important that there was sufficient pump-out time when switching precursors, otherwise cross-contamination occurred. Line pumping was typically performed overnight. Before $H_2O$ FEBIE, the dual beam chamber with the sample loaded underwent atmospheric plasma cleaning for 2 hours. The FEBIE pattern was defined as a single pixel line of length 1.5 $\mu$m that was positioned in the gap region, between the FEBID deposits. The flow rate was set to raise the chamber pressure ~5 x $10^{-5}$ mbar and the patterning time was 3 minutes. This procedure was used in all instances (Figures 1 - 3).

Ion milling was then performed using $O^+$ as the primary ion species. The beam energy and current were 30 kV and 30 pA, respectively. The dwell time was 100 ns and the pixel overlap was 95% in both vertical and horizontal scan directions (X and Y). The FIB mill pattern was defined as a 3 x 3 $\mu$m square with a masked region in the center, which did not undergo FIB milling. The was done using the subtractive boolean operator for overlaying patterns. This masked region was defined along the outer edges of the C FEBID deposits. In the case of rectangular prisms the approximate dimensions were 1 x 1 $\mu$m (Figure 1). For the circular dimer structures this consisted of 2 merged circle patterns of diameter 500 nm. In addition to this a single pixel wide line pattern was defined and positioned in the gap region between the C EBID boxes with equivalent dwell time and overlap (in both X and Y directions). Milling of the 2 defined patterns was performed in a parallel fashion, the total milling time was 190 s (Figure 1). In Figure 2 only a line pattern was defined, the mill time was ~10 s. Finally, the sample was removed from the microscope and treated in $O_2$ plasma (Glow Research) to remove the C mask. Plasma treatment was performed in 20 s pulses with rest times of 30 s to avoid sustained sample heating. This was repeated approximately 10 times.

**Raman:** Samples were analysed using a Renishaw Raman InVia Qontor Microscope (Renishaw plc., Wotton-under-Edge, UK), equipped with an air-cooled charge-coupled device (CCD). The spectrometer is fitted with holographic notch filters and a 2400 l/mm line visible spectrum grating. The attached microscope is a Leica DMLM equipped with a four standard objectives and a trinocular viewer that accommodates a video camera allowing direct viewing of the sample. The spectrometer was controlled by PC using the instrument control software (Renishaw WiRE$^{TM}$, Version 5.3). Before data collection the instrument was calibrated using a silicon internal standard.

A 24.5 $\mu$M solution was prepared by dissolving crystal violet (*Merck*) in ethanol. The solution was dropcast onto the sample and air dried before analysis. A reference spectrum of the crystal violet was collected by drop casting a 2 mM solution onto clean stainless steel. The acquisition parameters were identical. Sample excitation was achieved using a Renishaw RL532C CW diode laser (Renishaw plc., Wotton-under-Edge, UK) emitting 532 nm. Spectra were recorded using a × 100/0.85 NA objective over the spectral range of 600 - 1800 cm$^{-1}$ with the accumulation of 10 scans, with 1 s exposure and a laser power of 0.15 mW. Spectra were not corrected for instrument response.

**NiPS$_3$ / MnPSe$_3$ Nanoantennas** Thin films were mechanically exfoliated from commercial single crystals onto Si/ SiO$_2$(290 nm) substrates. Suitable flakes were identified in an optical microscope, and their thickness confirmed by tilted SEM imaging or AFM. The sample was then loaded into a plasma FIB dual beam system (Themofisher G4 Hydra Plasma FIB-SEM) and pumped to base pressure. Regions of interest were identified using the SEM with the gas injection system, (Multichem$^{TH}$ device pre-aligned to a distance of 200 $\mu$m above the sample surface. The patterning electron beam energy and current were 20 keV and 400 pA, respectively. To start a circle pattern was defined with diameter 350 nm, the total irradiation time was 240s resulting in a charge density of 6.23 x $10^{20}$ C/$\mu m^2$. An array was then built by changing the diameters from 200 nm to 600 nm in step sizes of 50 nm, maintaining the same electron dose per unit area. FIB milling was performed using a 30 kV, 30 pA Xe$^+$ beam. A 3 x 3 $\mu$m square pattern was defined with a circle cut-out in the center, the circle circumference was defined along the FEBID deposit edge. FIB milling was stopped when the underlying Si/SiO$_2$ surface was reached using end point detection.

Dark field spectroscopy was performed on a Nikon LV150N microscope with a fiber- coupled output. Sample illumination was achieved using a tungsten halogen lamp, a circular beam block with a diameter smaller than the beam diameter and a 50× Nikon (0.8 NA) dark-field objective. The objective only illuminated the sample at large angles to the nor- mal. Reflected light from the sample was guided back through the same objective toward a fiber coupler. Due to the small diameter of the multimode fiber core used, only light reflected back at small angles to the normal was collected. The fiber from the microscope was subsequently coupled to a Princeton Instruments spectrometer and charge coupled device. Finite-difference time-domain simulations were carried out using Lumerical Inc. software.

# Acknowledgement

J.S. and S.P. acknowledge the National Health and Medical Research Council (NHMRC) 2022 Ideas Grants, application ID 2021208 for funding support. We also acknowledge the facilities and the scientific and technical assistance of Sydney Analytical, Sydney Microscopy & Microanalysis and Research and Prototype Foundry, based at the Sydney Nanoscience Hub, core research facilities at The University of Sydney. L.S. acknowledges funding sup- port through a Humboldt Research Fellowship from the Alexander von Humboldt Foundation. P.G.Z., Y.W. and A.I.T. acknowledge support from EPSRC grants EP/S030751/1, EP/V006975/1, EP/V007696/1 and EP/V026496/1. Y.W. and A.I.T. acknowledge support from UKRI fellowship TWIST-NANOSPEC EP/X02153X/1.

## Supporting Information Available

**S1**: Extended characterization of the carbon deposit masking workflow including carbon deposition by beam skirting, omitting step (ii) focused electron beam induced etching in the gap region and comparison of inert $Xe^+$ and reactive $O^+$ primary ion species. **S2**: Line edge roughness measurements of adjacent rectangular prism Au structures. **S3**: Super- resolution focused ion beam milling. **S4**: Helium ion microscope benchmarking including mill gap profile and material swelling. **S5**: Supporting Raman scattering measurements. **S6**: Supporting $MnPSe_3$ nanoantenna data. **S7**: Supporting $MnPSe_3$ nanoantenna data.